\documentclass[aps,prl,twocolumn,superscriptaddress]{revtex4}

\usepackage{graphicx}
\usepackage{amsmath}
\usepackage{amssymb}
\usepackage{bm}
\usepackage[version=4]{mhchem}
\usepackage{upgreek}
\usepackage{hyperref}
\usepackage{float}
\usepackage{siunitx}

\begin{document}
\title{Evolution of Fullerenes in Circumstellar Envelopes by Carbon Condensation: Insights from Reactive Molecular Dynamics Simulations}

\author{Zhisen Meng}
\affiliation{Laboratory for Relativistic Astrophysics, Department of Physics, Guangxi University, 530004, Nanning, China}

\author{Zhao Wang}
\email{zw@gxu.edu.cn}
\affiliation{Laboratory for Relativistic Astrophysics, Department of Physics, Guangxi University, 530004, Nanning, China}

\begin{abstract}
Fullerenes, including \ce{C60} and \ce{C70}, have been detected in various astronomical environments. Understanding how their structures evolve over time is essential for gaining insights into their life cycle and making further observations. To address this, we conducted reactive molecular dynamics simulations to investigate the evolution of fullerenes in the circumstellar envelopes surrounding carbon-rich asymptotic giant branch stars. Our simulations employed a bottom-up chemistry scheme, wherein fullerenes grow by absorbing and condensing small carbon-based molecules. The results revealed the formation of different structures through heterogeneous reactions based on hydrogen concentration, leading to the emergence of onion-like nanostructures or single-layer fullerenes. To examine the impact of these structural changes on the infrared emission characteristics of fullerenes, we performed quantum chemical calculations. The results indicate that as fullerenes grow larger, additional emission features are introduced in the infrared spectrum. Moreover, two-layered fullerenes show noticeable blue-shift or weakening effects on the bands associated with out-of-plane vibration modes.
\end{abstract}

\maketitle
\section{Introduction} 

Fullerenes, such as \ce{C60} and \ce{C70}, were discovered by \cite{Kroto1985} in a laser ablation experiment on graphite, and were then confirmed by spectroscopic studies \citep{Kratschmer1990the, Kratschmer1990solid, Taylor1990}. Although they are considered to be stable in harsh interstellar conditions, direct observational evidence of cosmic fullerenes was not obtained until 2010 \citep{Cami2010}. Fullerenes have since been reported to exist in various astronomical environments, including asymptotic giant branch (AGB) stars \citep{Gielen2011}, reflection nebulae \citep{Berne2013, Sellgren2010}, planetary nebulae \citep{Cami2010, Garcia-Hernandez2010}, young stellar objects \citep{Roberts2012}, photodissociation regions \citep{Castellanos2014}, and diffuse interstellar medium (ISM) \citep{Berne2017}. Fullerenes are vital to the life cycle of large carbonaceous molecules/particles in interstellar space \citep{Candian2018}. Understanding their formation and evolution is hence crucial not only for their further observation but also for comprehending the evolution of carbon-rich dust, which is intricately linked to the formation, life, and death of stars \citep{Otsuka2014}.

There are two primary mechanisms for the formation of fullerenes: top-down and bottom-up chemistry. Fullerenes have often been observed to coexist with polycyclic aromatic hydrocarbons (PAHs) \citep{Sellgren2010, Dunk2013, Garcia2013}, with abundance observed to decrease with increasing concentration of PAHs \citep{Berne2012}. This has led to the exploration of the top-down formation mechanism, in which large PAH molecules or other carbonaceous dust are transformed into fullerenes \citep{Chuvilin2010, Zhang2013, Pietrucci2014, Garcia-Hernandez2010, Micelotta2012}. Dehydrogenation of PAHs is considered a critical step in this process, occurring before the folding of a large PAH into a fullerene cage \citep{Montillaud2013, Berne2015, Patra2014}. 

On the other hand, the bottom-up approach, which is more commonly considered, is assumed to take place in the dense and hot envelopes of evolved stars. It typically involves reactions between carbon chain molecules followed by hydrocarbon addition \citep{Cherchneff2000, Pascoli2000, Bettens1997, Jones2011, Parker2012}. Based on ab initio calculations \citep{Bates1997}, it has been demonstrated that closed-shell molecules with adjacent pentagons exhibit thermodynamic stability. Subsequently, Morokuma, Irle, and coworkers introduced the concept of the ``shrinking hot giant road'' using density functional tight-binding molecular dynamics simulations \citep{Irle2003a, Irle2003b, Zheng2004, Zheng2005, Zheng2007, Irle2006, Irle2007, Saha2009}. This proposed mechanism suggests that the formation of larger fullerene molecules (comprising more than 100 atoms) initiates from carbon chains through self-catalytic reactions within the thermal carbon vapor. As these larger fullerenes are exposed to ultraviolet radiation, they undergo shrinkage reactions, transforming into \ce{C60} and \ce{C70}, while releasing smaller molecular fragments like \ce{C2} \citep{Irle2007, Berne2012, Berne2015}. 

Despite extensive efforts to elucidate the formation mechanisms of interstellar fullerenes, their subsequent evolution in ISM remains poorly understood. The scarcity of \ce{C60} detections in planetary nebulae suggests that these molecules may undergo evolutionary reactions and structural transformations in space, leading to significant changes in their spectral features \citep{Otsuka2014, Zhen2019}. Laboratory experiments are valuable for simulating interstellar reactions, but they often face challenges in determining the chemical pathway due to the presence of highly reactive intermediates with short lifetimes. Quantum chemical calculations (QCCs) provide a solution by assisting in the identification of intermediates and the elucidation of reaction pathways \citep{Parker2017, Fioroni2019}. However, the computational cost associated with QCCs often limits their applicability in studying large molecules such as fullerenes. In this context, classical molecular dynamics (MD) simulations based on reactive force fields offer a promising alternative \citep{Meuwly2019}. These simulations strike a balance between computational accuracy and efficiency, enabling the investigation of interstellar formation of carbon-based large molecules and nanoparticles under complex conditions \citep{Patra2014, Montillaud2014, Qi2018, Chen2018, Chen2020, Ostroumova2019, Hanine2020, Orekhov2020, Anders2021, Tang2022}.

In this study, we utilize reactive MD simulations to investigate the process of fullerene evolution in the circumstellar envelopes (CSEs) of carbon-rich AGB stars. The CSEs of AGB stars, including dust condensation zone, are well-known for their abundance of small carbonaceous chains such as \ce{C2}, making them primary sites for the production of interstellar carbonaceous dust, and earning them the designation of ``cosmic dust factories'' \citep{Clayton1995, Bakker1997, Millar2000, Santoro2020}. The main objective of this study is to provide valuable insights into the structural evolution of fullerenes, driven by the adsorption of carbonaceous chains, under various simulated conditions that mimic the CSEs of AGB stars.

%##########################################################################
\section{Methods} 

Surface adsorption reactions have been identified as a significant mechanism for the growth of interstellar dust \citep{Carpentier2012, Burke2015}. In the bottom-up chemistry approach, the growth of fullerenes is primarily driven by the adsorption and enrichment of small carbonaceous molecules. The dust condensation zone within the CSEs of carbon-rich AGB stars offers a favorable astronomical environment for these adsorption reactions, as it provides the necessary materials and thermal energy for the reactions to take place \citep{Decin2010, Carpentier2012}.

Given that the primary objective of this study was to explore the enrichment and evolution of \ce{C2} small molecules on the surface of stable \ce{C60} fullerene, we conducted simulations to investigate the condensation process of \ce{C2} molecules onto a \ce{C60} seed particle. The initial configuration consisted of a periodic simulation cell with dimensions of $80\times80\times80$ \AA$^3$, with the \ce{C60} molecule placed at its center. A total of 200 \ce{C2} molecules ($N_\text{C}=400$) were gradually introduced into the cell at random positions over a duration of $20$ ns. Additionally, a variable number of neutral hydrogen atoms $N_\text{H}$ were included in the system, with different \ce{H} concentrations corresponding to $N_\text{H}/N_\text{C}$ ratios of 0, 0.05, 0.15, or 0.50.

\begin{figure}
\centerline{\includegraphics[width=8.6cm]{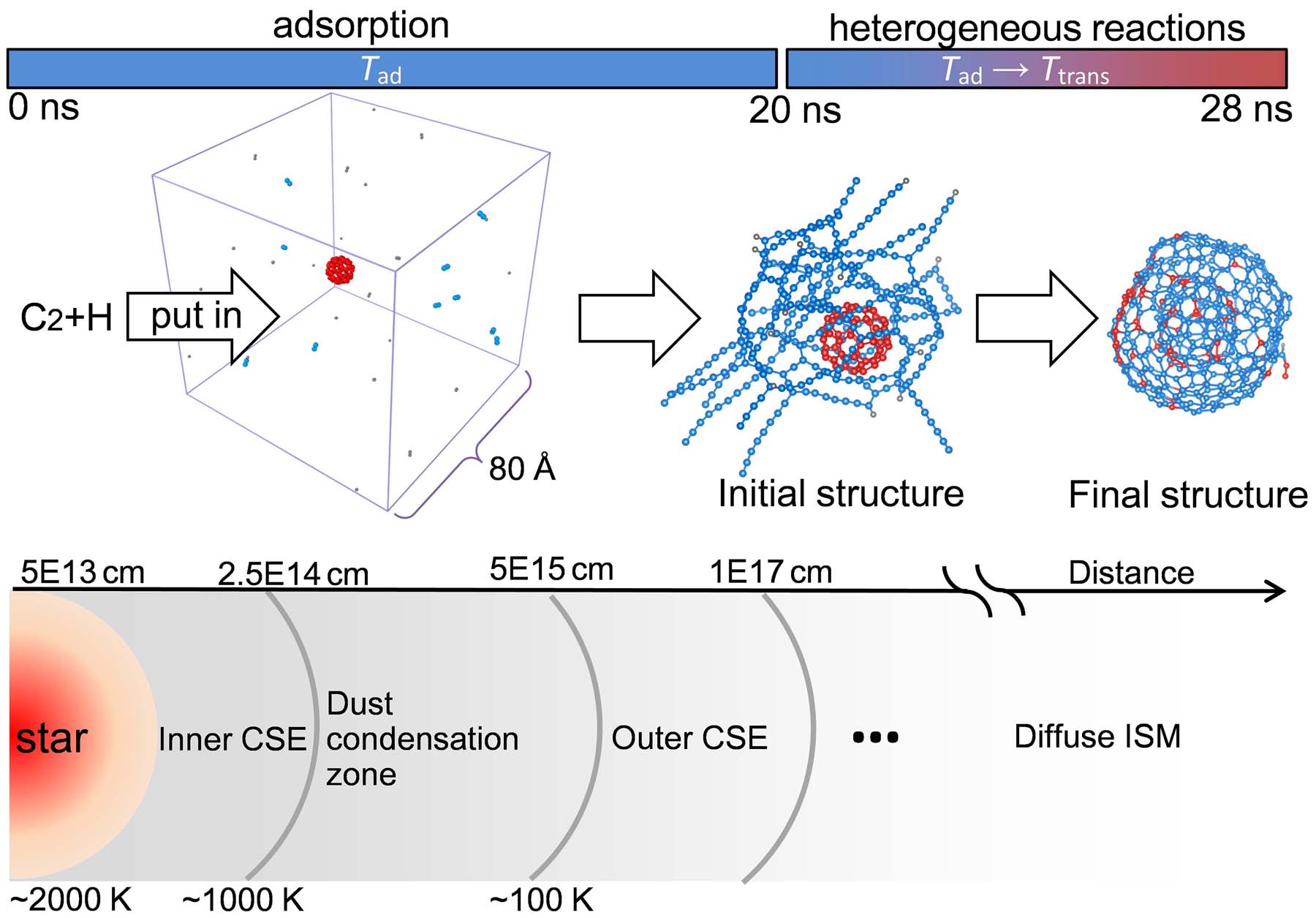}}
\caption{Schematics for the simulated evolution of \ce{C60} via adsorption reactions.}
\label{F1}
\end{figure}

During the simulations, the majority of the added \ce{C2} and \ce{H} particles adsorbed onto the surface of the \ce{C60} seed particle through van der Waals interactions \citep{Wang2019, Wang2023}. This adsorption process facilitated chemical reactions between the \ce{C2} and \ce{H} particles, leading to the formation of an initial structure as depicted in Figure \ref{F1}. The simulations were conducted at a specified temperature $T_{\text{ad}}$ to mimic the thermal energy available in the dust condensation zone within the CSEs of carbon-rich AGB stars. The thermal energy in this region provides favorable conditions for the adsorption reactions to take place and contributes to the growth of fullerenes through the enrichment of carbonaceous molecules.

To simulate the evolution of the formed initial structure, the temperature was progressively increased from $T_{\text{ad}}$ to $T_{\text{trans}}$ over a duration of $8$ ns. This temperature increase led to the transformation of the initial structure into a final structure, whose characteristics depended on the specific temperature values and the hydrogen concentration. These factors are known to be influenced by the condition of the CSE and the distance from the central AGB star \citep{Nowotny2005, Freytag2017}. Although the representative temperature near AGB star regions is $2000$ to $3500$ K \citep{Ziurys2006}, the dust formation regions in its CSEs are typically found at distances ranging from $2.5\times10^{14}$ to $5\times10^{15}$ cm from the star's center, with temperatures estimated to be in the range of $100$ to $1000$ K \citep{Decin2010}. Therefore, our choice of $T_{\text{ad}}$ fell within this temperature range. However, we set $T_{\text{trans}}$ to be in the range of $1500$ to $4000$ K in order to accelerate the reactions, as the actual timescale of the reactions is much longer than what can be simulated using MD. Hence, the specific value of $T_{\text{trans}}$ does not hold direct thermodynamic significance but serves as a parameter to represent the reaction rate within the constraints of the simulation time.

The interactions between atoms in our simulations were modeled using the ReaxFF reactive force field \citep{Ashraf2017}. ReaxFF is an interatomic potential function that takes into account bond-order dependence and consists of eight terms: $E_{\text{bond}}, E_{\text{over}}, E_{\text{under}}, E_{\text{lp}}, E_{\text{val}}, E_{\text{tors}}, E_{\text{vdW}},$ and $E_{\text{Coulomb}}$, as shown in Equation \ref{eq1}. 

\begin{equation}
\label{eq1}
E_{\text{p}}=E_{\text{bond}}+E_{\text{over}}+E_{\text{under}}+E_{\text{lp}}+E_{\text{val}}+E_{\text{tors}}+E_{\text{vdW}}+E_{\text{Coulomb}}.
\end{equation}

\noindent These terms describe the potential energy of the system with respect to bond length and order, overcoordination penalty, undercoordination stability, lone pair, bond angle, single bond torsion, van der Waals interactions, and Coulomb interactions, respectively. The detailed expression of each term, parameterization, and benchmarking of the ReaxFF force field can be found in \citet{Ashraf2017}. We chose the ReaxFF force field for its proven success in modeling the formation of fullerenes and related molecules or nanostructures using MD simulations, as demonstrated in previous studies such as \citet{Martin2017, Mainitz2016, Mao2017, Ostroumova2019, Orekhov2020, Li2022}. 

In this study, the simulations were performed using the Large-scale Atomic/Molecular Massively Parallel Simulator (LAMMPS) \citep{Plimpton1995}. The temperature control was achieved using the canonical Nos\'e-Hoover thermostat with a time step of $0.1$ fs. It is important to note that, for simplicity, we considered \ce{C2} molecules as the initial reactants, although other carbon sources, such as larger carbonaceous chains or PAHs, could also be present \citep{Millar2000, Li2008, Li2019, Agundez2020, Kovacs2020, Meng2021, Mcguire2022, Meng2023}. The ratio between the number of C and H atoms, $N_\text{H}$/$N_\text{C}$, was determined based on the abundance ratio of $\leq 0.03$ between \ce{C2} and \ce{C60} in IRC+10216 \citep{Clayton1995}, while the number of hydrogen atoms was set to balance the formation and photodissociation of \ce{C-H} bonds \citep{Patra2014}. In the Supplementary Data\footnote{https://github.com/mengzss/Fullerene\_Evolution.git}, we have included an example of the simulation code used in our study, as well as the corresponding simulation outputs. This will facilitate readers who are interested in replicating the simulations and conducting further investigations. 

%########################################################################
\section{Results and discussions} 
\subsection{Initial Structures}

\begin{figure}
\centerline{\includegraphics[width=8.5cm]{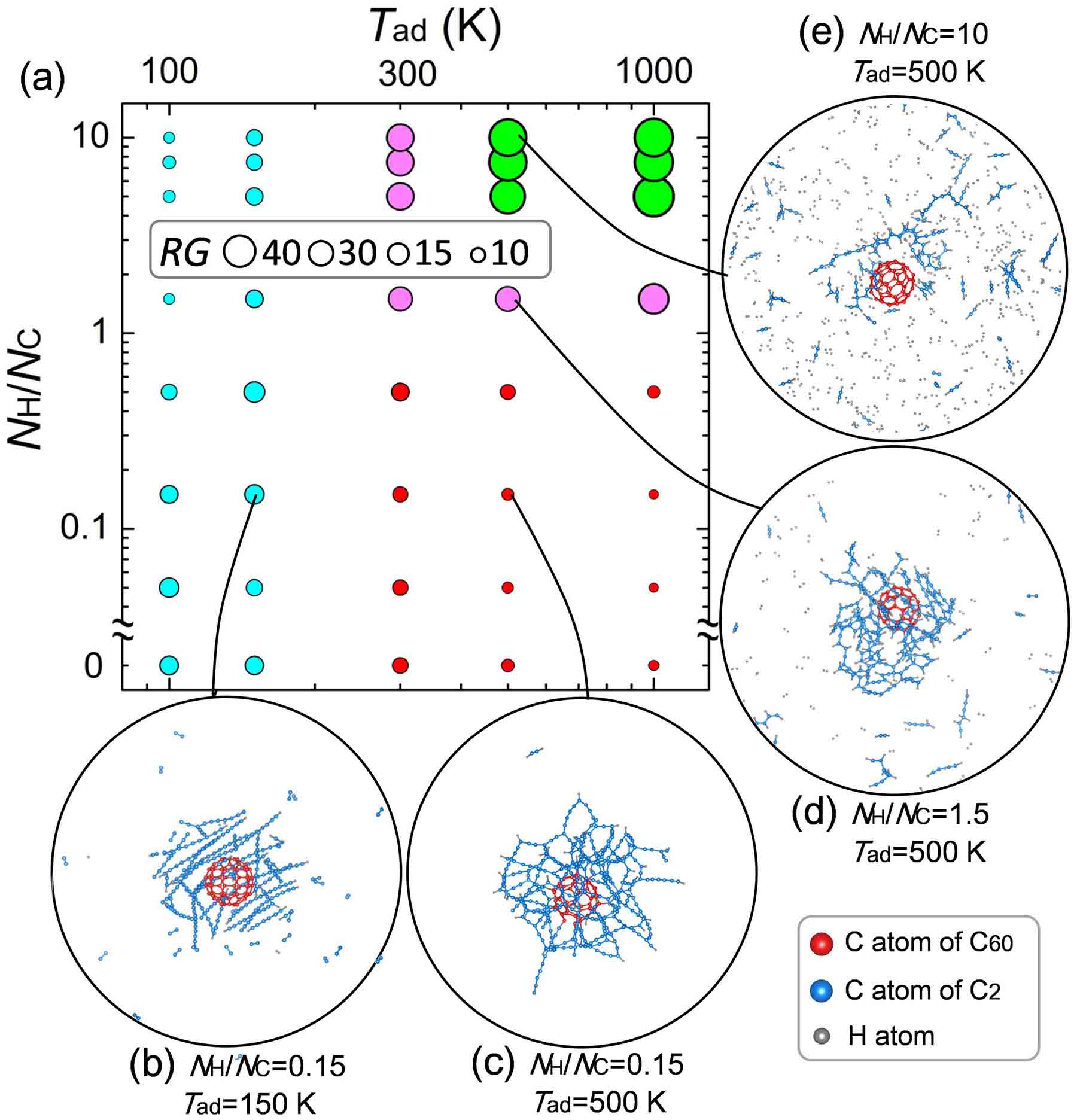}}
\caption{(a) $RG$ of the initial structures formed with different $N_\text{H}$/$N_\text{C}$ at various adsorption temperatures $T_\text{ad}$. The bubbles' size and colors represent the $RG$ value and different classes of the initial structures, respectively. (b-e) Snapshots of the formed initial structures.}
\label{F2}
\end{figure}

In our simulations, we observed the formation of different initial structures with varying shapes under different conditions, as depicted in Figure \ref{F2}. The compactness of these structures was characterized using the radius of gyration $RG$, which is a measure of the distribution of atoms from the center of mass \citep{Lobanov2008}. A lower $RG$ indicates a more condensed structure for a given number of atoms. In Panel (a) of Figure \ref{F2}, the size of the circles represents the $RG$ values of the initial structures. We observed that at low concentrations of H, the initial structures exhibited a high degree of condensation, as indicated by the small circles in the bottom of Panel (a). In these cases, the adsorbed \ce{C2} molecules formed aligned long carbon chains at low temperatures, as shown in Panel (b). At higher temperatures, we observed that the chains interconnected to form cage-like structures, as shown in Panel (c). This is in agreement with the observation of \citet{Anders2021}, which suggests that warm organic cluster particles have a tendency to stick together through chemical reaction when colliding.

On the contrary, at high \ce{H} concentrations, it becomes challenging for the \ce{C2} molecules to form the same types of structures. This is due to many \ce{C2} molecules being saturated with \ce{H} atoms, which occupy the chemically active sites and hinder the formation of interconnected chains. This observation aligns with the findings of \citet{Cherchneff2011}, who observed from the molecular abundance of IRC+10216 that excess hydrogen led to the formation of hydrocarbon molecules rather than carbon chains. As a result, the \ce{C2} form small hydrocarbon molecules such as \ce{C2H2}, \ce{C2H4}, \ce{C4H6}, and \ce{C5H5}, which remain as individual entities. These molecules are primarily held together by vdW forces on the surface of the \ce{C60} at low temperatures. However, when the temperature exceeds the binding energy between these molecules, they can no longer remain bound, leading to their dissociation, as shown in Panel (e) of Figure \ref{F2}. The case depicted in Panel (d) represents an intermediate scenario between the condensed structure and the dissociated state. It is worth noting that previous studies have demonstrated that excessive hydrogenation can lead to the dissociative fragmentation of molecules \citep{Tang2022, Gatchell2015}.

\subsection{Final Structures}

\begin{figure}
\centerline{\includegraphics[width=9cm]{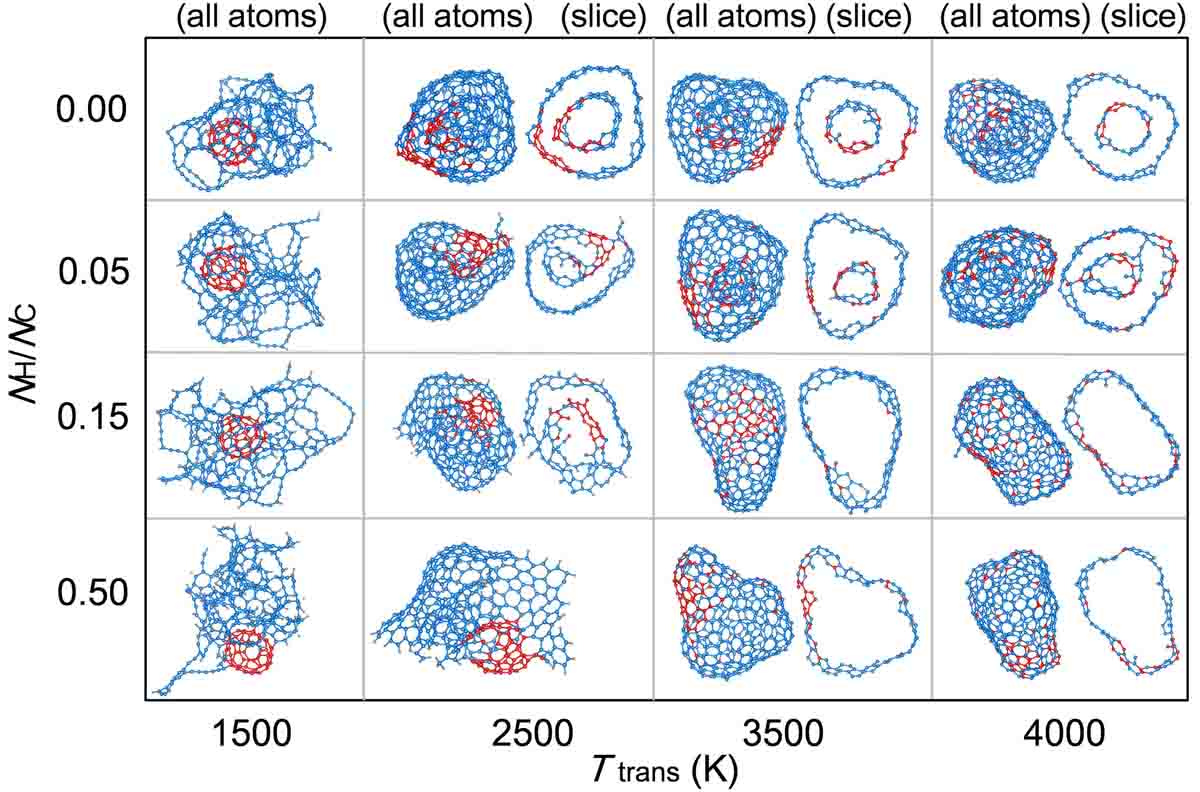}}
\caption{Final structures formed with different H concentrations through heterogeneous reactions at different temperatures. The gray, red, and blue dots represent H and C atoms originating from \ce{C60} and \ce{C2}, respectively.}
\label{F3}
\end{figure}

We explore further possible evolution of the condensed initial structures. In the CSEs of AGB stars, the dust particles experience various astronomical events such as photoprocessing, stellar winds, and shocks, which could result in a sharp increase in their temperature and induce heterogeneous reactions \citep{Jones2013}. In our simulations, after the initial structures were formed through adsorption, we subjected them to high temperatures to explore their potential structural transformations. Our results revealed two distinct evolution pathways leading to different types of final structures. In the case of low \ce{H} concentration, onion-like nanostructures consisting of two layers of fullerenes were observed, as depicted in the top panels of Figure \ref{F3}. Conversely, for the case of high \ce{H} concentration, single-layer fullerenes were formed, as shown in the bottom panels of Figure \ref{F3}. 

It is worth noting that the specific temperature at which the transformation occurs does not definitively determine the type of final structure, except in the case of $2500$ K, which appears insufficient to trigger the formation of fullerene within the simulated time scale. This finding is consistent with the observation of \citet{Cami2018}, which shows that PAHs become the main product of evolution in the hydrogen-rich environment at temperatures between $1000$ and $1700$ K. Moreover, it is important to acknowledge that fullerene transformation could still potentially occur at that temperature, taking into account that the actual reaction time is significantly longer than what is in the MD study despite of a much lower concentration.

\begin{figure}
\centerline{\includegraphics[width=9cm]{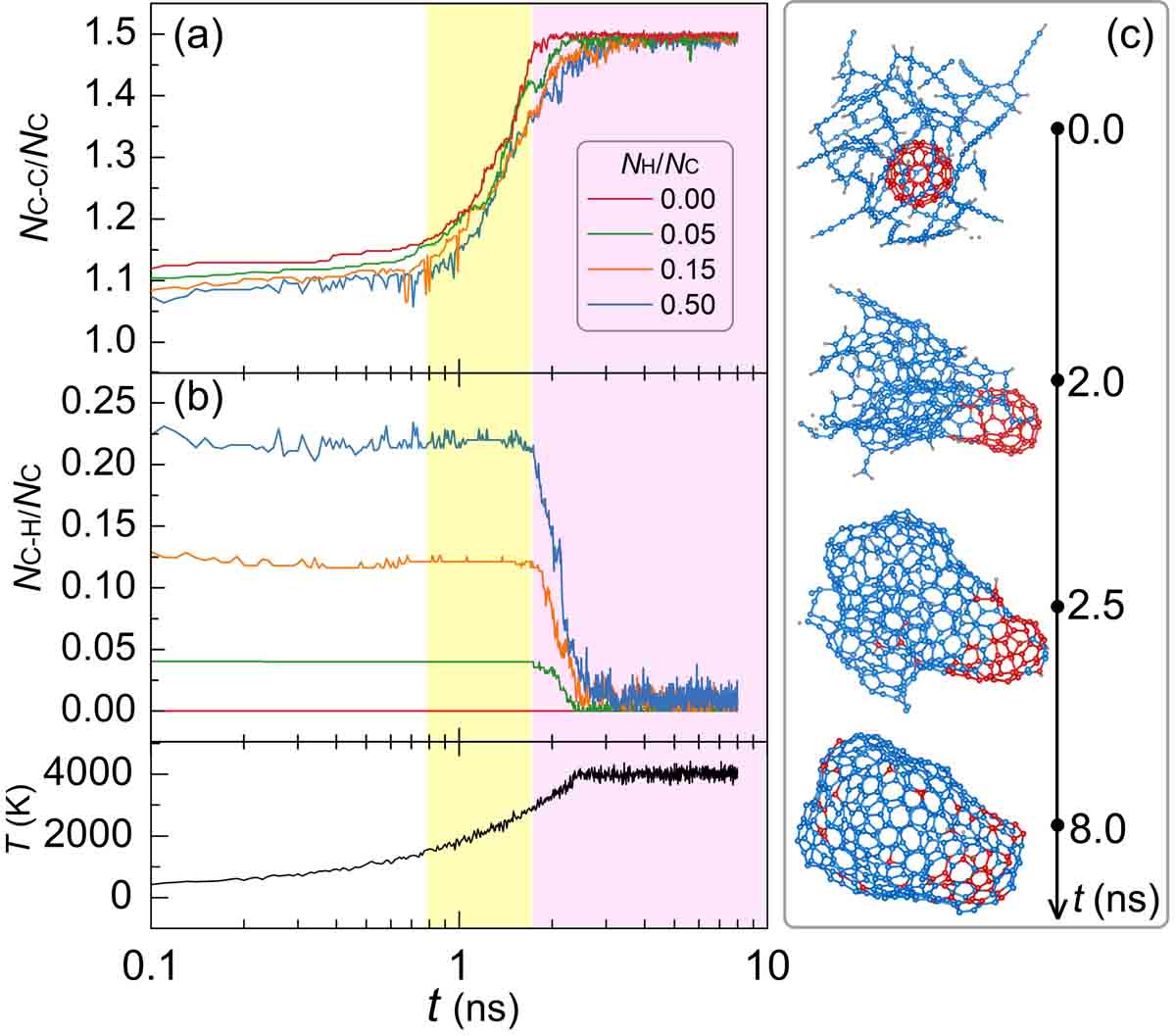}}
\caption{Evolution of \ce{C-C} (a) and \ce{C-H} bonds (b) during heating to $T_\text{trans}=4000$ K for different H concentrations. The background colors indicate three stages of structural transformation. (c) Configurations of the simulated system at different time for the case of $N_\text{H}/N_\text{C}=0.5$.}
\label{F4}
\end{figure}

The \ce{C-H} bonds played a key role in determining the final structures. The plot in Figure \ref{F4} illustrates the evolution of the number of \ce{C-C} and \ce{C-H} bonds, denoted as $N_\text{\ce{C-C}}$ and $N_\text{\ce{C-H}}$, revealing three distinct stages of structural transformation. Initially, during the first $0.8$ nanoseconds, the bond numbers remain relatively stable. In the second stage, a notable increase in $N_\text{\ce{C-C}}$ occurs as the temperature surpasses a specific threshold ($1500$ K), indicating a significant structural transition. This transition involves the connection of adsorbed molecules through unoccupied sites, with the highest bonding rate observed in the absence of hydrogen. This process is similar to the ring-condensation process outlined in \cite{Irle2003a}, marking the phase in which carbon rings commence formation. Subsequently, at around $1.5$ nanoseconds and at a temperature of approximately $2500$ K, a substantial breakage of \ce{C-H} bonds commences. This leads to the generation of new unoccupied carbon sites available for bonding, which typically promotes the formation of new carbon rings \citep{Saha2012}. 

The selection between the two distinct types of final structures predominantly occurred during the transition from the second to the third stage. During this stage of evolution, the predominant process involves cage closure, characterized by the collapse of a sizable ring that undergoes a closing mechanism \citep{Irle2003a}, while the closure of a double-layered fullerene was not observed. In scenarios with low hydrogen concentration, the majority of adsorbed carbon atoms formed bonds with each other in the second stage, resulting in the creation of a relatively stable layer that enveloped the central \ce{C60}. The temperature during this phase was insufficient to facilitate the transformation of intra-layer $sp^{2}$ bonds into inter-layer $sp^{3}$ bonds. Conversely, in high hydrogen concentration scenarios, the breakage of \ce{C-H} bonds in the third stage led to the emergence of numerous unoccupied carbon sites. As the temperature increased in the third stage, these sites interacted with the central \ce{C60}, eventually causing it to dissolve into a larger, single-layered fullerene cage. This process is depicted in Panel (c) of Figure \ref{F4} taking the case of $T_\text{trans}=4000$ K for example. 

During the process of compact cage contraction, we observed events of small molecule pop-out, involving species like \ce{C2} and \ce{H2}, referred to as the ``shrinking hot giant'' road in \cite{Irle2007}. It is worth noting that in cases of low hydrogen concentration, when the temperature was high enough, the \ce{C60} could also react with atoms in the outer layer; however, the two-layered structures were ultimately maintained. While direct evidence is lacking, it is plausible that dissociated hydrogen atoms played a catalytic role in the opening of the central \ce{C60} \citep{Yang2023}. We note that the formation of multilayer fullerenes studied here differs from the previous high-density pure carbon gas ultrafast cooling mechanism \citep{Ostroumova2019}. The adsorption-driven process studied here is more likely to occur in the long-term evolution of low-density interstellar environments.

\begin{figure}
\centerline{\includegraphics[width=8.0cm]{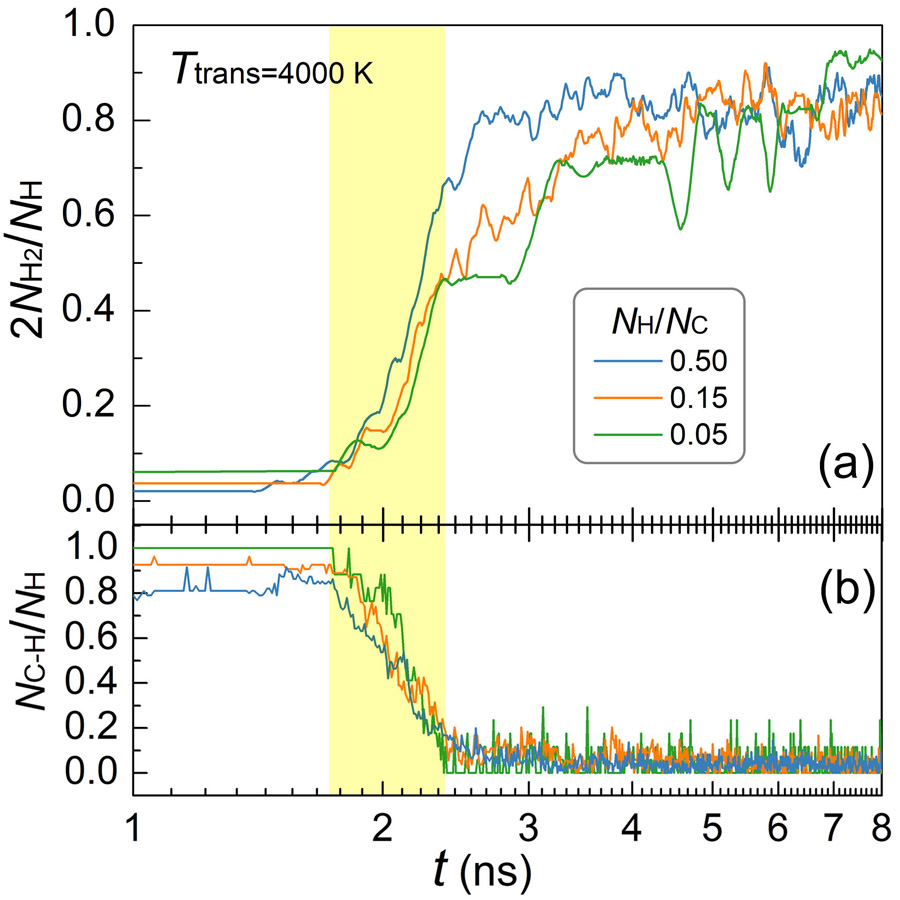}}
\caption{(a) Ratio of the number of \ce{H2} molecules formed over the total number of H atoms initially present in the simulation cell as a function of time, for different H concentrations. (b) Time evolution of the \ce{C-H} bonds in the system.}
\label{F5}
\end{figure}

The formation of \ce{H2} molecules in the ISM remains a challenging question in astronomy \citep{Wakelam2017}. The exothermic nature of the combination of hydrogen atoms to form \ce{H2} poses a dilemma, as \ce{H2} lacks an efficient mechanism to dissipate the released heat in the low-density environment of the ISM. Current understanding suggests that the primary route for interstellar \ce{H2} formation is through the Eley-Rideal reaction on dust or ice particles, which act as heat reservoirs \citep{Bauschlicher1998, Mennella2012}. These particles play a crucial role in \ce{H2} formation by facilitating the dissipation of heat \citep{Gould1963, Lemaire2010, Thrower2012, Mennella2012, Boschman2015, Foley2018, Barrales-martinez2019, Pantaleone2021}. Conversely, hydrogen can also facilitate the formation of complex organic molecules \citep{Lu2021, Yang2023}. In the simulations conducted for this study, the formation of \ce{H2} molecules was observed as by-products during the growth of fullerenes. It was found that the formation of \ce{H2} is closely correlated with the breaking of \ce{C-H} bonds, as shown in Figure \ref{F5}, suggesting a potential catalytic role of evolving fullerenes in the \ce{H2} formation process. However, it is important to note that the specific question regarding heat dissipation during \ce{H2} formation was not directly addressed in these simulations, as a global thermostat was applied. To fully understand the intricacies of heat dissipation in the \ce{H2} formation process, a separate set of simulations specifically designed to explore this aspect and more in-depth analyses would be necessary.

\subsection{Infrared spectrum}

\begin{figure}
\centerline{\includegraphics[width=8.5cm]{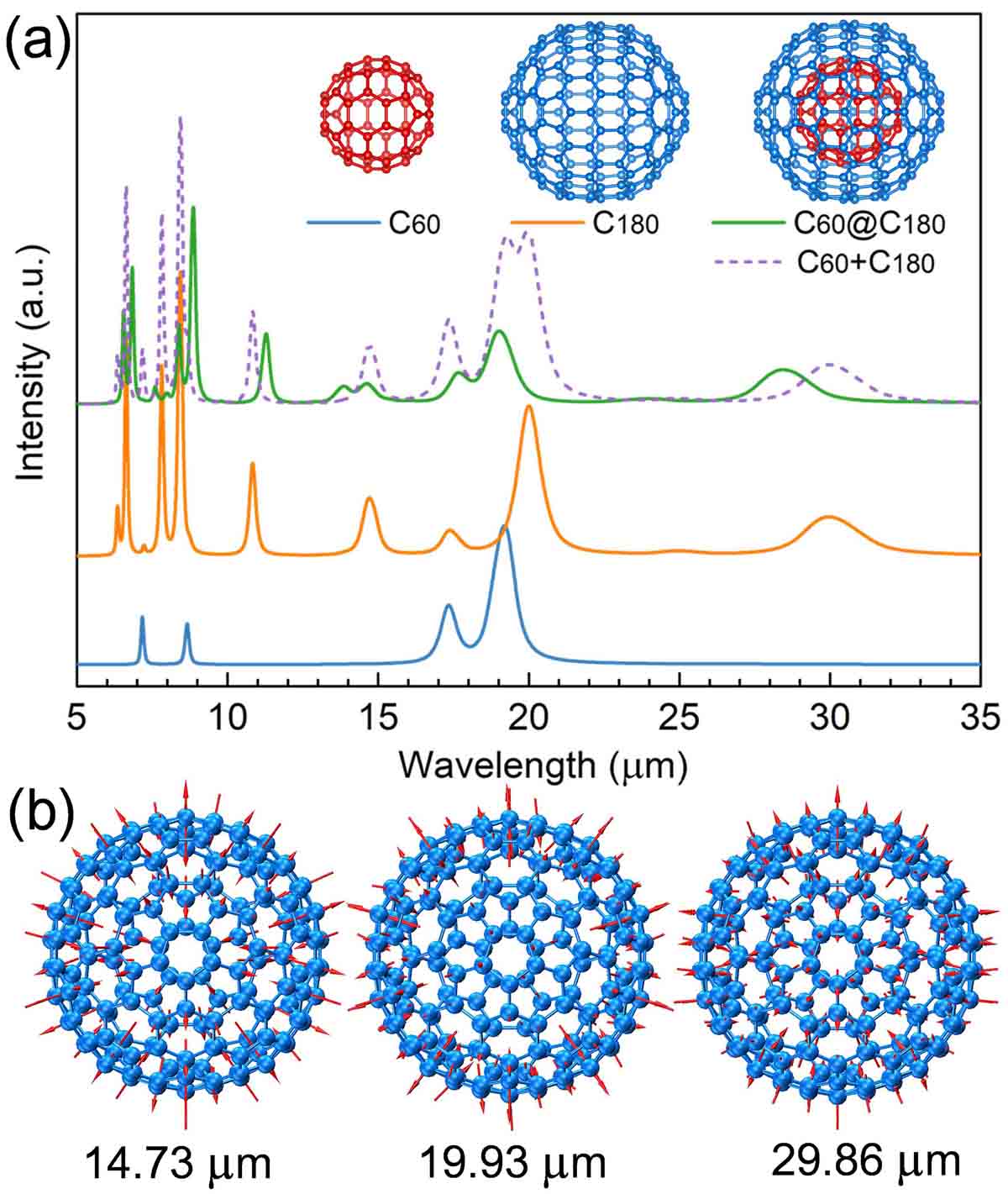}}
\caption{(a) Calculated  IR spectrum of \ce{C60}, \ce{C180}, and \ce{C60}@\ce{C180}. The dashed line represents the sum of the blue and orange lines. (b) Visualizations of the vibration modes of \ce{C180} at 14.73, 19.93, and 29.86$\,\upmu$m. The arrows indicate the vibration directions of atoms.} 
\label{F6}
\end{figure}

To explore the impact of the evolution towards the two different types of fullerenes on their infrared spectra, we employed density functional theory (DFT) for geometry optimizations and frequency calculations \citep{Banhatti2021}. The calculations were conducted at the B3LYP/6-31G(d) level of theory for three selected samples: a \ce{C60} molecule, a \ce{C180} molecule, and a two-layered \ce{C60}@\ce{C180} composite, as depicted in the inset of Figure \ref{F6}. We chose these smaller molecules for simplicity, considering the significant computational challenges associated with performing calculations on larger simulated final structures. Given the large size of the system, we computed the spectra under the assumption of harmonic vibrations. To account for anharmonicity effects, we uniformly scaled the calculated spectra by a factor of 0.9613, as proposed by \citet{Borowski2012}. This scaling factor enables us to approximate the influence of anharmonicity on the spectra and improve their accuracy. Additionally, for an accurate description of the interaction between different layers of fullerenes \citep{Saal2009}, we included the D3 version of Grimme's empirical dispersion interaction \citep{Grimme2010}, along with the Becke-Johnson damping function \citep{Grimme2011}.

Figure \ref{F6} (a) displays the computed IR spectra of the three samples. It is noted that when comparing \ce{C180} with \ce{C60}, additional emission features are observed at specific wavelengths ($10.84$, $14.65$, and $29.86\,\upmu$m) for \ce{C180}. The complex vibrational behavior of \ce{C180} indicated by the additional features aligned with previous observation of \cite{Bakowies1991, Cami2010}. Upon comparing the summed spectrum of \ce{C60} and \ce{C180} (purple, dashed line) with that of \ce{C60}@\ce{C180} (green line), notable observations are made. Specifically, the $14.73$, $19.93$, and $29.86\,\upmu$m bands of the summed spectrum exhibit hindrance or a blue-shifted pattern when the two molecules are combined. This alteration in the IR features, characterized by a decrease in intensity accompanied by an increase in frequency, can typically be attributed to the increased rigidity of the molecule in the direction of vibration when the \ce{C60} molecule is encapsulated within \ce{C180}. This suggests that these bands likely correspond to vibrations perpendicular to the surface. A more detailed analysis of the vibrational modes associated with these bands, as depicted in Figure \ref{F6} (b), provides further confirmation of this hypothesis. Therefore, the blue-shifted behavior of these bands holds particular interest for astronomical observations targeting onion-like fullerenes.

%##########################################################################
\section{Conclusions} 

In conclusion, this study employed MD simulations to investigate the evolution of fullerenes in environments resembling those found in the CSEs of AGB stars. The simulations revealed two distinct types of transformations that occur when fullerenes adsorb carbon chains, resulting in the formation of single- and double-layered carbon nanostructures. The formation of single-layered structures was observed under high H concentration, while the formation of double-layered structures occurred under low H concentration. We note that the simulations did not investigate the transformation of fullerenes into larger sizes with additional layers due to system size limitations. Additionally, the study found that the rate of \ce{H2} molecule formation is closely linked to the breaking of \ce{C-H} bonds in the evolving fullerene, implying a potential catalytic role of fullerene in \ce{H2} formation.

Furthermore, DFT calculations were conducted on three fullerene molecules to investigate the impact of structural evolution on IR emission features. The results demonstrated that as fullerenes undergo structural evolution towards larger sizes, additional emission features emerge at specific wavelengths, such as $10.84$, $14.65$, and $29.86\,\upmu$m. Additionally, the presence of a double layer in the fullerene structure led to noticeable blue-shift or weakening effects on the bands at $14.73$, $19.93$, and $29.86\,\upmu$m. These findings provide valuable insights into the structural changes and spectral characteristics of fullerenes, suggesting the possibility of fullerene growth in the ISM through the adsorption of small carbon molecules and subsequent heterogeneous reactions.

\section*{Acknowledgments}
The authors acknowledge financial support from the National Natural Science Foundation of China (11964002). Prof. Yong Zhang is acknowledged for fruitful discussion.

\section*{Data Availability}
The Supplementary Data include an example of the LAMMPS input script that was used in our study, along with the corresponding simulation outputs. They are available at \href{https://github.com/mengzss/Fullerene\_Evolution.git}{https://github.com/mengzss/Fullerene\_Evolution.git}. 

%\bibliography{refs}

\end{document}